\documentstyle[aps,pre,twocolumn,epsfig,floats]{revtex}

\newcommand{\be}{\begin{equation}\FL}
\newcommand{\ee}{\end{equation}}
\newcommand{\beas}{\begin{eqnarray*}}
\newcommand{\eeas}{\end{eqnarray*}}
\newcommand{\bea}{\begin{eqnarray}\FL}
\newcommand{\eea}{\end{eqnarray}}
\newcommand{\req}[1]{(\ref{#1})}

\begin{document}
\draft
\wideabs{
\title{Nonequilibrium phase transition in a model for social influence. }
\author{Claudio Castellano$^{(1)}$, Matteo Marsili$^{(2)}$ 
and Alessandro Vespignani$^{(3)}$}

\address{
$^{(1)}$ Fachbereich Physik, Universit\"at GH Essen
D-45117 Essen, Germany\\
$^{(2)}$Istituto Nazionale per la Fisica della Materia (INFM),
Trieste-SISSA Unit, V. Beirut 2-4, I-34014 Trieste \\
$^{(3)}$The Abdus Salam International Centre for Theoretical Physics, 
P.O. Box 586, I-34014 Trieste}

\date{\today}
\maketitle

\begin{abstract}
We present extensive numerical simulations of the Axelrod's model for 
{\em social influence}, aimed at understanding the formation
of cultural domains. This is a nonequilibrium model with short 
range interactions and a remarkably rich dynamical behavior. 
We study the phase diagram of the model and uncover a nonequilibrium
phase transition separating an ordered (culturally polarized) phase
from a disordered (culturally fragmented) one. The nature of 
the phase transition can be continuous or discontinuous depending 
on the model parameters. At the transition, the size of cultural regions
is power-law distributed.
\end{abstract}

\pacs{PACS numbers: 87.23.Ge, 05.40.-a, 05.70.Ln}
}

Recently, the study of complex systems entered social science in order to
understand how self-organization, cooperative effects and adaptation
arise in social systems\cite{axel1}. In this context the use of simple
automata or dynamical models often elucidates the mechanisms at the
basis of the observed complex behaviors\cite{axel1,Anderson88}.

In this spirit, R. Axelrod has recently proposed an interesting model
to mimic how dissemination of culture works\cite{axel97,axel96}.
Culture is used here to indicate the set of individual
attributes, such as language, art or technical standards
subject to {\em social influence}, i.e. that can be changed as effect
of mutual interactions.
The automaton does not consider the effect of central institutions
or mass media and focuses on the self-organization resulting from
a simple local dynamics representing the social influence.
This dynamics is assumed to satisfy two simple properties:
i) individuals are more likely to interact with others who already share
many of their cultural attributes; ii) interaction
increases the number of features that individuals share.
Starting from an initial state with features distributed randomly
this leads to the formation and coarsening of regions of shared culture.

In this Letter we carry out an accurate numerical analysis of 
Axelrod's model that unravels a remarkably rich behavior, not
detected in previous investigations.
Depending on the initial degree of disorder, the model undergoes
a phase transition separating an ordered from a disordered phase.
The ordered phase is characterized by the growth of a dominant cultural
region spanning a large fraction of the whole system.
On the contrary, in the disordered phase the system freezes in a highly
fragmented state with a nontrivial distribution of the sizes of cultural
regions.
Such a fragmented configuration is reached in a finite time,
which diverges at the phase transition.
In the whole ordered phase instead, the coarsening process lasts
for a time proportional to the system size, before freezing
into the culturally polarized state.
Interestingly, the nature of the transition turns from continuous to
discontinuous when the number of cultural features is increased.
Close to the transition, the distribution of region sizes follows
a power law in agreement with the results of Ref.
\cite{gomes99} on the statistical analysis of the diversity of
languages.
Some of these features are captured by a mean field approximation that we
discuss below.

Axelrod's model is defined on a square lattice of linear size $L$. 
On each site $i$ there is a set 
of $F$ integer variables $\sigma_{i,f}$ which define the cultural
``features'' of the individuals living on that site. 
In the original model, each feature $f=1,\ldots,F$ on each site 
$i$ is initially drawn randomly from 
a uniform distribution on the integers between $1$ and $q$.
The parameter $q$ is a measure of the initial cultural variability
(i.e. disorder) in the system. 
Here, we relax the constraint of integer $q$ by using a 
Poisson distribution ${\rm Prob}(\sigma_{i,f}=k)=q^k e^{-q}/k!$, that 
allows $q$ to be a positive real number. 
Though the results are qualitatively the same, our choice is more
convenient to study the behavior of the model as $q$ varies.

At each time step, a pair of nearest neighbor sites
$i$ and $j$ is randomly chosen. A feature $f$ is chosen
randomly and if $\sigma_{i,f}\not =\sigma_{j,f}$ nothing happens.
If instead $\sigma_{i,f}=\sigma_{j,f}$ then an additional feature
$f'$ is randomly chosen among those taking different values across the
bond, $\sigma_{i,f'}\neq \sigma_{j,f'}$.
Such a feature is then set equal: $\sigma_{i,f'}\to
\sigma_{i,f'}'=\sigma_{j,f'}$.
A ``sweep'' of the lattice, i.e. $L^2$ such time steps,
defines the time unit as usual.
Axelrod's model (at least in the original formulation) can be seen
as $F$ coupled voter models~\cite{Liggett85}.

During the dynamical evolution the total diversity, measured as the number
of different values of a feature $f$ which are present in the system,
always decreases.
Clearly if all features are equal across a bond
($\sigma_{i,f}=\sigma_{j,f}$, $\forall f$) or if they are all
different ($\sigma_{i,f}\not =\sigma_{j,f}$ $\forall f$) no change 
can occur on the bond $ij$.
A configuration such that, on each bond, either all
features are equal or they are all different is an {\em absorbing state}:
Dynamics will stop if such a state is reached.
There are clearly many absorbing states:
The dynamics on any finite lattice converges to one of them.
The final state can be characterized  by the distribution of cultural
region sizes, where a region is defined as the connected set of sites
sharing exactly the same features.

The dynamical evolution is characterized by the competition
between the disorder of the initial configuration and 
the ordering drive due to the local social interactions. 
It is intuitively clear that when $q$ is small the initial state is almost
completely uniform, whereas for large values of $q$ almost all sites have
features $\sigma_{i,f}$ totally different from those of their nearest
neighbors.
In the two cases we expect the system to converge to a uniform or a highly
fragmented state, in which interaction or disorder dominate, respectively.
In order to understand how these two limit situations are connected as\
$q$ varies, we have simulated the Axelrod's model for a
number of features ranging  from $F=2$ to $10$ and sizes up to $L=150$.
We first discuss the dependence of the final state on the parameters
$q$ and $F$ and then the dynamical behavior of the model.

\begin{figure}
\centerline{\epsfig{figure=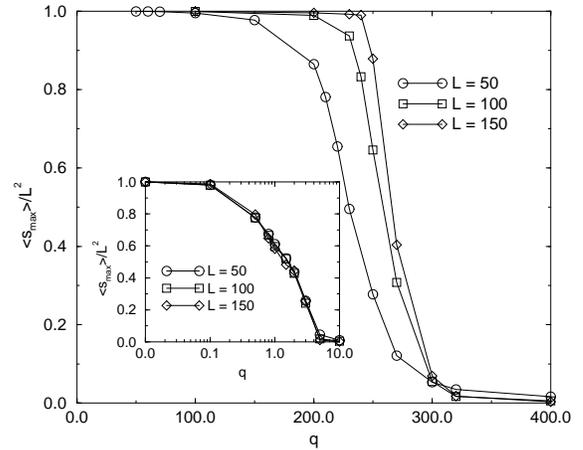,width=7cm,angle=-90}}
\caption{Behavior of $s_{\max}/L^2$ vs $q$ for three different system sizes
and $F=10$. In the inset the same quantity is reported for $F=2$.}
\label{Fig1}
\end{figure}

{\em The frozen state}. In any finite lattice the
dynamics converges to a frozen absorbing state. The existence
of a transition in the properties of the final absorbing states
is very clear from the plot (Fig.~\ref{Fig1})
of the average size of the largest region $\langle s_{\max}\rangle$ as a
function of $q$ for $F=10$: For $q =q_c\approx 300$ we observe a sharp
transition characterized by a sudden drop of $\langle s_{\max}\rangle/L^2$
which becomes steeper and steeper for increasing sizes $L$.
This points to the existence of a transition between a ``culturally
polarized phase'' for $q < q_c$, where one of the regions has a size
of the order of the whole system,
and a ``culturally fragmented'' phase, where all domains are finite.
The transition is of the first order, with
the size of the largest region having a finite  discontinuity at $q=q_c$.
Note also that, for $q<q_c$ the largest domain approaches a unitary density,
i.e. it invades the whole system.
This scenario holds for all values $F>2$ investigated.

\begin{figure}
\centerline{\epsfig{figure=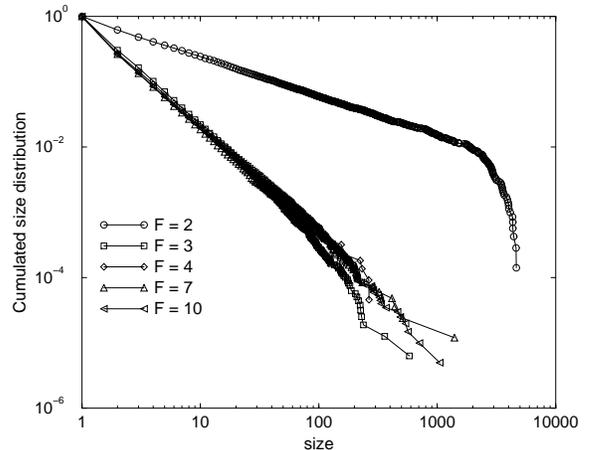,width=7cm,angle=-90}}
\caption{Cumulated distribution of region sizes for $q \approx q_c$,
$L=100$ and several values of $F$.}
\label{Fig2}
\end{figure}

The situation is different for $F=2$ (inset of Fig.~\ref{Fig1}): The
fraction occupied by the largest cluster $\langle
s_{\max}\rangle/L^2$ vanishes continuously as $q\to q_c^-$.
The difference between $F=2$ and $F>2$ is confirmed and clarified
by the study of the size distribution of cultural regions at the transition.
Let $P_L(s,q)$ be the probability distribution of regions of $s$
sites in a system of size $L$. The cumulated distribution is plotted
in Fig.~\ref{Fig2} for several $F$ and values of $q$ around the transition.
Fig.~\ref{Fig2} shows that $P_L(s,q)$ decays as a power law
$s^{-\tau}$ and that the exponent $\tau$ is universal
($\tau \simeq 2.6$) for $F>2$ but takes a different value for $F=2$
($\tau \simeq 1.6$).
In particular we find $\tau_{F>2}> 2$ and $\tau_{F=2}<2$.

The different nature of the transition for $F$ equal or greater than 2
can be related to the exponent $\tau$.
Let $N(q,L)$ be the total number of regions in the system.
Requiring the total area to be $L^2$ leads to
\be
L^2=N(q,L)\langle s\rangle=N(q,L)\sum_{s=1}^{\infty} s\, P_L(s,q).
\label{moments}
\ee
For $q>q_c$ there are $N(q,L)\sim L^2$ domains of finite size and the
sum on $s$ is finite as $L\to\infty$.
On the other hand, for $q<q_c$, there are few small domains and a
large one of size $s_{\max}\sim L^2$. Hence the probability distribution
can be written in the generic scaling form
$P_L(s,q)=s^{-\tau} {\cal F}(s/s_{co})+A(q)\delta_{s,s_{\max}}$
where $s_{co}$ is a cut-off scale, the function ${\cal F}(x)$ 
is constant for $x \ll 1$ and decays very rapidly for $x \gg 1$
and $A(q)=0$ for $q>q_c$.

The divergence $L^2$ in the l.h.s. of Eq.~\req{moments} is matched,
for $q<q_c$, by the component $A\delta_{s,s_{\max}}$ in $P_L(s,q)$,
with $s_{\max}\sim L^2$.
The nature of the transition is identified by the behavior
of $A(q)$ for $q \to q_c^-$.
For $\tau<2$, similarly to what happens in percolation theory
\cite{percolation}, the transition occurs through the divergence of
$s_{co}$ and hence of a correlation length, as $q\to q_c^+$.
This causes the divergence of $\langle s\rangle$ in  Eq.~\req{moments}
(because $s P_L(s,q) \sim s^{1-\tau}$ with $\tau<2$), which matches the
divergence of the l.h.s. of Eq.~(\ref{moments}) as $L\to\infty$,
through usual finite size scaling arguments \cite{percolation}.
Indeed, at $q=q_c$ the cut-off diverges with the system size as
$s_{co} \sim L^D$.
On the other side of the transition a
similar divergence of the cutoff $s_{co}$ occurs
and the amplitude $A$ must vanish as $q\to q_c^-$.
This scenario is typical of second order phase transitions and agrees
with the value $\tau<2$ found for $F=2$ and Fig.~\ref{Fig1}.

When $\tau>2$ the previous scenario cannot hold: For $q> q_c$,
even if $s_{co}$ diverges, the sum on $s$ in Eq.~\req{moments}
remains finite. Hence as $q\to q_c^+$ the number of
domains $N(q,L)$ must remain of order $L^2$.
On the other side of the transition, the $L^2$ term
in the l.h.s. of Eq. \req{moments} can only be matched by the term
$s_{\max}\sim L^2$. Since no divergence arises from the sum on small
components as $q\to q_c^-$, the amplitude $A$ of the large component
must remain of order one in this limit.
We then conclude that, as $q$ crosses $q_c$, the
nature of the distribution changes abruptly for $\tau>2$.
In particular the amplitude $A(q)$ exhibits a discontinuous jump across
the transition~\cite{notebose}.

\begin{figure}
\centerline{\epsfig{figure=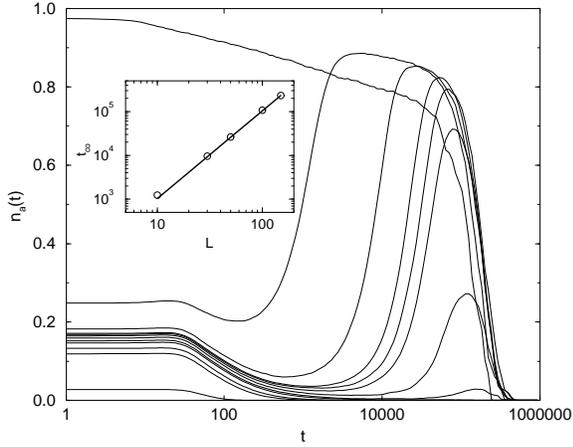,width=7cm,angle=-90}}
\caption{$n_a(t)$ for $F=10$,
$q=1$, 100, 200, 230, 240, 250, 270, 300, 320, 400, 500, 10000
(top to bottom), $L=150$.
The inset reports the dependence of the freezing time $t_{co}$ on
$L$ for $F=10$ and $q=100<q_c$. The bold line has slope 2.}
\label{Fig3}
\end{figure}

{\em Dynamics}:
In order to investigate the model dynamics we study
the density $n_a$ of active bonds. An active bond is a bond across
which at least one feature is different and at least one is equal, so
that there can be some dynamics. This quantity is indicative of the
dynamical state of the system, being zero in frozen configurations.
In Fig.~\ref{Fig3} we show the behavior of $n_a$ as a function
of time for different values of $q$ and $F=10$.
For large values of $q$, after a short initial transient, 
the density of active bonds decays rapidly and the system locks into a frozen
configuration in a finite time.
For small $q$ instead, $n_a(t)$ displays a slow decay with a
large majority of bonds in the active state until it falls abruptly for
long times. 
As the system size is increased, the slow decay extends to longer
and longer times: The cut-off time, at which activity suddenly dies
scales as $t_{co} \sim L^2$ as shown in the inset of Fig.~\ref{Fig3}.
For a large range of intermediate $q$ values,
$n_a$ first decreases almost to zero but then rises again towards a
peak of activity, from which the slow decay begins.

The dynamics is essentially a coarsening process of homogeneous regions.
When the process lasts only for a finite time, as for $q>q_c$,
it gives rise to regions of finite size.
On the contrary, for $q<q_c$ the coarsening process goes on for a
time $t_{co}\sim L^2$  and produces a region of size comparable with
the whole system.
The exponent $z=2$ relating $t_{co}$ to $L$ is the same found
in nonconserved phase-ordering~\cite{Bray94} and in particular in
the two-dimensional voter model~\cite{Scheucher88}, which also exhibits
a slow decay of the density of active bonds~\cite{Frachebourg96}.

For an infinite system ($L\to\infty$) and $q<q_c$ the
system is indefinitely in a coarsening state.
Therefore we can also define the transition as separating two
different dynamical regimes, with a fast decay of $n_a$ lasting for
a finite characteristic time for $q>q_c$ and a slow, infinitely long decay
for $q<q_c$.
For $F=2$ the behavior is qualitatively the same: Noticeably, 
no evident signature of the different nature of the transition
(continuous vs. discontinuous) can be inferred from the dynamical evolution.

The dynamical behavior of the model can be studied within a 
single bond  mean field treatment. Let $P_m(t)$ be the probability that a
randomly picked bond is of type $m$ at time $t$, i. e.  $m$
features across the bond are equal and $F-m$ are different.
At $t=0$, since features are assigned uncorrelated random values, we have
$P_m(0)={F\choose m} \rho^m_0 (1-\rho_0)^{F-m}$ where
$\rho_0={\rm Prob}[\sigma_{i,f}=\sigma_{j,f}]$
is the probability that two sites have feature $f$ with the same value
at $t=0$
In the mean field approximation, $P_m$ satisfies the master equation
\bea
\frac{d P_m}{dt}&=&\sum_{k=1}^{F-1}\frac{k}{F}P_k
\left[\delta_{m,k+1}-\delta_{m,k}+\right.\nonumber\\
&~&\left.(g-1)\sum_{n=0}^F\left(P_n
W_{n,m}^{(k)}-P_mW_{m,n}^{(k)}\right)\right],
\label{mastereq}
\eea
where $g$ is the lattice coordination number and
$W_{n,m}^{(k)}$ is the transition probability from a $n$-type bond
to a $m$-type bond due to the updating of a $k$-type neighbor bond.
This equation describes how the number of bonds of type $m$ is
affected by the dynamics: $P_k\, k/F$ is the probability to select a
bond of type $k$ and one of the $k$ features which are equal across it.
If $k=m-1$ a new bond of type $m$ is created
whereas if $k=m$ one bond of type $m$ is deleted.
This explains the first two terms on the r. h. s. of Eq.~(\ref{mastereq}).
But the change of a feature of the site may also affect the state
of the other $g-1$ bonds connecting such site to its neighbors.
The remaining terms in Eq.~(\ref{mastereq}) take into account this
possibility.
We can compute the probabilities 
$W_{n,m}^{(k)}$ by analyzing in detail each possible process. 
Let us consider e.g. a type $k=1$ bond adjacent to a type $n=0$ one.
Without loss of generality we can let the features on the extreme sites
of the $1$ bond be $(0,0,\ldots,0)$ and $(0,\sigma_2,\ldots,\sigma_F)$,
with $\sigma_j\neq 0$. The latter site is shared also by the $0$ bond
whose other site has features
$(\sigma_1',\sigma_2',\ldots,\sigma_F')$ with
$\sigma_1'\neq 0$ and
$\sigma_j'\neq \sigma_j$ for $j>1$. 
When the $1$ bond becomes a type $2$ bond,
$\sigma_2\to 0$ the neighbor $0$ bond can either remain a $0$ bond,
if $\sigma_2'\neq 0$ or it can become a $1$ bond if $\sigma_2'=0$. In
the spirit of the mean field  approximation we introduce the
probability
$\rho={\rm Prob}(\sigma_2'=0)$ which now becomes time dependent, as
we shall see. Then $W_{0,0}^{(1)}=1-\rho$ and $W_{0,1}^{(1)}=\rho$.
In much the same way we can compute the other transition rates. For
example, if $F=2$, the only non zero elements are
\be
\begin{array}{lll}
W_{0,0}^{(1)}=1-\rho, & W_{0,1}^{(1)}=\rho& ~\\
W_{1,0}^{(1)}=1/2, & W_{1,1}^{(1)}=(1-\rho)/2, &
W_{1,2}^{(1)}=\rho/2\\ W_{2,1}^{(1)}=1.&~&~
\end{array}
\ee

\begin{figure}
\centerline{\epsfig{figure=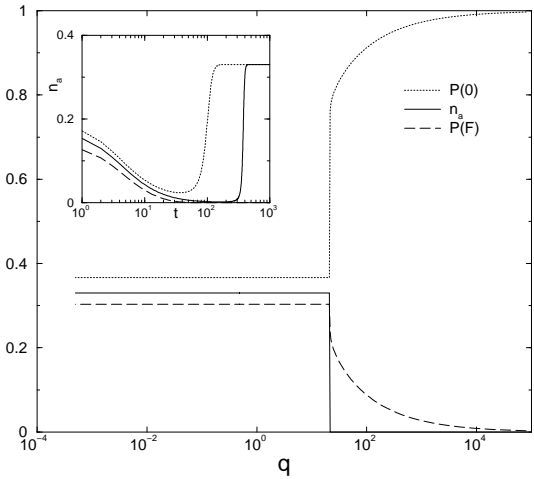,width=7cm}}
\caption{Phase diagram within the mean field approximation for $F=3$.
In the inset the mean field behavior of $n_a(t)$ for $q=15<q_c$ (dotted),
$q=20\simeq q_c$ (full) and $q=25>q_c$ (dashed).
In the mean field description the coarsening phase lasts indefinitely because 
it is implicitly assumed an infinite size system.}
\label{Fig4}
\end{figure}

The system of equations \req{mastereq} is closed once the dynamics
of $\rho(t)$ is specified. The simplest such equation, in the
mean-field spirit, is $\rho=\sum_k k P_k/F$. This amounts to
say that between any two sites there is a bond, and hence that the
probability that a  feature across that bond takes the same value can
be expressed in terms of the $P_m$\cite{note1}.
The numerical integration of the mean field equations yields the 
phase diagram of the model as shown in Fig.~\ref{Fig4}, which exhibits
a phase transition.
The order parameter $n_a=\sum_{k=1}^{F-1} P_k$ undergoes a
discontinuity at $q=q_c$, as it jumps from a
finite value for $q<q_c$ to zero for $q>q_c$.
The mean field fairly reproduces also the dynamical evolution of the model
as can be seen from the inset of Fig.~\ref{Fig4}.

Despite the extreme simplicity of Axelrod's model, it is tempting 
to compare the picture derived in this Letter with some recent 
analysis on the diversity of languages\cite{gomes99}, which to some 
extent can be considered as an indicator of cultural homogeneity.
Gomes and coauthors, by analyzing the statistics of more 
than 6700 languages, have found that the 
number ${\cal L}$ of linguistically homogeneous regions
inhabited by $N$ individuals scales as 
${\cal L}\sim N^{-\tau}$, with $\tau\simeq 1.5$ for populations smaller
than $6 \times 10^6$ individuals. For larger populations the exponent
changes suggesting a different dynamical evolution.
These findings can be compatible with our continuous phase transition 
scenario ($\tau\simeq 1.6$) assuming the equivalence between language 
and cultural features.
However it remains unclear why language spreading is
self-organized close to the transition point or why the value $F=2$
should be relevant for the process.

In summary, we have presented numerical simulations of the Axelrod's
model for social influence. Social interaction that tends to
make culture homogeneous competes with the disorder
introduced by the number of different traits each cultural feature
initially has.
We find a critical value separating two phases in which one of
the above elements (interaction or disorder) respectively dominates. 
The nature of the phase transition and the emergent collective behavior 
is analyzed by common tools of statistical physics. 
We find that the transition is continuous or discontinuous depending
on the model parameters. 
Our study shows that the use of concepts and methods developed
in physics may be of help in the context of social sciences.
For instance, the prevalence, for small values of $q$, of one of
the initially equivalent cultures can be regarded as a spontaneous
symmetry breaking due to stochastic fluctuations.
It would be interesting, in the future, to analyze extensions
of the simple Axelrod's model to take into account additional ingredients
like the presence of migrations or the effect of 
geographical barriers.

C. C. acknowledges support from the Alexander Von Humboldt foundation.
A. V. acknowledges partial support from the European Network contract
ERBFMRXCT980183.


\begin{thebibliography}{99}
\bibitem{note1}
	Qualitatively, the results we obtained  do not depend on
	the equation for $\rho$, which can indeed even be considered as a 
	time independent constant.
\end{thebibliography}
\end{document}